\documentclass{article}
\usepackage{amssymb, amsmath}
\usepackage{amsfonts}
\usepackage{commath}
\usepackage{graphicx}
\usepackage{pdfpages}
\usepackage{multirow}
\usepackage[ruled,vlined]{algorithm2e}
\author{
  Changye Wu\thanks{CEREMADE, Universit\'e Paris-Dauphine, France. wu@ceremade.dauphine.fr}
  \and
  Christian Robert\thanks{Universit\'e Paris Dauphine PSL, CREST, France and University of Warwick, UK.  xian@ceremade.dauphine.fr}
}
\title{Average of Recentered Parallel MCMC for Big Data}
\begin{document}
\maketitle
\begin{abstract}
\noindent In big data context, traditional MCMC methods, such as Metropolis-Hastings algorithms and hybrid Monte Carlo, scale poorly because of their need to evaluate the likelihood over the whole data set at each iteration. In order to rescue MCMC methods, numerous approaches belonging to two categories: divide-and-conquer and subsampling, are proposed. In this article, we study parallel MCMC techniques and propose a new combination method in the divide-and-conquer framework. Compared with some parallel MCMC methods, such as consensus Monte Carlo, Weierstrass Sampler, instead of sampling from subposteriors, our method runs MCMC on rescaled subposteriors, but shares the same computation cost in the parallel stage. We also give a mathematical justification of our method and show its performance in several models.  Besides, even though our new method is proposed in parametric framework, it can been applied to non-parametric cases without difficulty. 
\end{abstract}
\section{Introduction}
\noindent  Due to the massive influx of data, the power of traditional MCMC algorithms is inhibited for Bayesian inference, for MCMC algorithms are difficult to scale. Indeed, MCMC algorithms, such as Metropolis-Hastings (MH) algorithms (\cite{Robert}), require at each iteration to sweep over the whole data set, which is very expensive on large data sets. In order to overcome this shortcoming and rescue MCMC algorithms for big data, a lot of efforts have been devoted over recent years to develop scalable MCMC algorithms. These approaches can be classified into two classes (\cite{Angelino}, \cite{Bardenet et al}): divide-and-conquer approaches (\cite{EPasAW},\cite{Mposterior},\cite{Neiswanger},\cite{WASP},\cite{ScottCMC},\cite{Wang}) and subsampling approaches(\cite{Bardenet2}, \cite{Chen}, \cite{Korattikara}, \cite{Welling}). In this article, we propose a new method belonging to the divide-and-conquer category. Specifically, we divide the whole data set into batches and repeat each batch a certain amount of times, run MCMC over repeated batches, recenter all subposteriors thus obtained and take their average as an approximation of the true posterior.\\
\\
Our article extends the traditional parallel MCMC algorithms in three directions. First, we scale each likelihood of the subposterior with a factor such that it could be regarded as an approximation of the true likelihood, by which we mean turning each subposterior covariance matrix into the same scale with that of the true posterior. Second, our combination method is simple enough, has solid mathematical justifications and is efficient. Third, even though our method is justified in parametric framework, it can be extend to non-parametric Bayesian without modification.\\
\\
The organization of this paper is as follows. Section 2 outlines the methodology and provides a mathematical justification of its validation. Section 3 applies the method to four numerical experiments and shows its power and efficiency. Section 4 concludes and discusses further research. 
\section{Averaging and Recentering Subposterior Distributions}
\noindent Let $\mathcal{X} = \{X_1, \cdots, X_N\}$ denote the data set and suppose that $X_i$ be i.i.d. observations from a common distribution $P_\theta$ possessing a density $f(x|\theta)$ where $\theta\in\Theta$, an open set of $\mathbb{R}^d$. We fix $\theta_0\in\Theta$, which may be regarded as the "true value" of the parameter.  Suppose the whole data set $\mathcal{X}$ be divided into $K$ subsets $\mathcal{X}_1, \cdots, \mathcal{X}_K$ with same size $M = N/K$. Denote 
\begin{equation*}
\ell(\theta, x) = \log{f(x|\theta)}
\end{equation*}
\begin{equation*}
\ell_i(\theta) = \sum_{j = 1}^M \log{f(x_{ij}|\theta)}
\end{equation*} 
\begin{equation*}
L_N(\theta) = \sum_{i=1}^K\sum_{j=1}^M\log{f(x_{ij}|\theta)} = \sum_{i=1}^K \ell_i(\theta)
\end{equation*}
\begin{equation*}
\hat{\theta}_i = \arg\max_{\theta\in\Theta}\ell_i(\theta), \quad \hat{\theta} = \arg\max_{\theta\in\Theta}L_N(\theta), \quad \bar{\theta} = \frac{1}{K}\sum_{i=1}^K\hat{\theta}_i
\end{equation*}
for each $i\in\{1,\cdots,K\}$, $\mathcal{X}_i = \{x_{ij}\}_{j=1}^M$. In classical parallel approaches, one decomposes the overall posterior into a product of subposteriors:
\begin{equation*}
\pi(\theta|\mathcal{X})\propto \prod_{i=1}^K\left(\pi(\theta)^{1/K}\exp(\ell_i(\theta))\right)
\end{equation*}
Even though this decomposition is correct mathematically, it is not reasonable in statistics. In Bayesian analysis, the type of prior should not change with the size of data set. Hence, using a prior that depends on the observation size is not appropriate. In order to overcome this shortcoming, we can create an artificial data set for each subset, which just repeats each data point $K$ times for each subset. Hence, we can apply the overall prior on these artificial data sets. That is, we regard the following rescaled subposteriors as approximations to the overall posterior:
\begin{equation*}
\pi_i(\theta|\mathcal{X}_i) \propto \exp\{K\ell_i(\theta)\}\pi(\theta)
\end{equation*}
This idea has also appeared in  \cite{Mposterior}, \cite{WASP}. Denote 
\begin{equation*}
\theta^*_i = \mathbb{E}_{\pi_i}(\theta), \quad \bar{\theta}^* = \frac{1}{K}\sum_{i=1}^K\theta^*_i
\end{equation*}
In these approximations, the factor $K$ rescales the variance of each subset posterior $\pi_i(\theta|\mathcal{X}_i)$ to be roughly of the same order as that of the overall posterior $\pi(\theta|\mathcal{X})\propto \exp(L_N(\theta))\pi(\theta)$. Inspired by this phenomenon, we recenter each subset posterior to their common mean and then average them to approximate the true posterior. That is, the overall posterior $\pi(\theta|\mathcal{X})$ is approximated by 
\begin{equation*}
\frac{1}{K}\sum_{i=1}^K\pi_i(\theta-\bar{\theta}^*+\theta^*_i|\mathcal{X}_i)
\end{equation*}
\begin{algorithm}[H]
\SetAlgoLined
\caption{Average of Recentered Subposterior\label{IR}}
\SetKwInOut{Input}{Input}\SetKwInOut{Output}{Output}
\Input{$K$ subsets of data $\mathcal{X}_1, \cdots, \mathcal{X}_K$, each with size $M$. }
\Output{Samples to approximate the true posterior.}
\For {$i = 1$ \KwTo $K$ (in parallel)}{
    \For {$t = 1$ \KwTo $T$}{
    Draw $\theta_i^t$ from $\pi_i(\theta|\mathcal{X}_i)$ via MCMC.
    }
    Calculate $\theta^*_i = \frac{1}{T}\sum_{t=1}^T\theta^t_i$.
}
Calculate $\bar{\theta}^* = \frac{1}{K}\sum_{i=1}^K\theta^*_i$\;
\For {$i = 1$ \KwTo $K$ (in parallel)}{
    \For {$t = 1$ \KwTo $T$}{
    $\theta_i^t \leftarrow \theta_i^t - \theta_i^* + \bar{\theta}^*$.
    }
}
\KwResult{$\{\theta_i^t| i=1,\cdots; K, t = 1,\cdots,T\}$ approximates the overall posterior.}
\end{algorithm}      
\noindent In order to proceed the theoretical analysis, we make some mild assumptions on the likelihood and the prior $\pi(\theta)$. These assumption are standard for the Bernstein-von Mises theorem (\cite{Ghosh}).\\ 
\\
\textbf{Assumption 1:} The support set $\{x: f(x|\theta) > 0\}$ is the same for all $\theta \in \Theta$.\\
\\
\textbf{Assumption 2:} $\ell(\theta,x)$ is three times differentiable with respect to $\theta$ in a neighbourhood $\{\theta: ||\theta-\theta_0||\leq \delta_0\}$ of $\theta_0$. The expectation of $\mathbb{E}_{\theta_0}\triangledown\ell(\theta_0, X_1)$ and $\mathbb{E}_{\theta_0}\triangledown^2\ell(\theta_0, X_1)$ are both finite and for any $x$ and $p, q, r \in \{1, \cdots, d\}$, 
\begin{equation*}
\sup_{\theta: ||\theta-\theta_0||\leq \delta_0}\frac{\partial^3}{\partial \theta_p\partial\theta_q\partial\theta_r}\ell(\theta,x)\leq M(x) \quad \text{and} \quad \mathbb{E}_{\theta_0}M(X_1)< \infty
\end{equation*}
\textbf{Assumption 3:} Interchange of the order of integrating with respect to $P_{\theta_0}$ and differentiation at $\theta_0$ is justified, so that
\begin{equation*}
\mathbb{E}_{\theta_0}\triangledown\ell(\theta_0, X_1)=0\quad \text{and}\quad \mathbb{E}_{\theta_0}\triangledown^2\ell(\theta_0, X_1) = - \mathbb{E}_{\theta_0}\triangledown\ell(\theta_0, X_1)[\triangledown\ell(\theta_0, X_1)]^T
\end{equation*}
Also the Fisher information $ I(\theta_0) = \mathbb{E}_{\theta_0}\triangledown\ell(\theta_0, X_1)[\triangledown\ell(\theta_0, X_1)]^T $ is positive definitely. \\
\\
\textbf{Assumption 4:} For any $\delta>0$, there exists an $\epsilon >0$, with $P_{\theta_0}-$probability one, such that
\begin{equation*}
\sup_{\theta:||\theta-\theta_0||>\delta}\frac{1}{N}(L_N(\theta)-L_N(\theta_0))<-\epsilon
\end{equation*}
for all sufficiently large $N$.\\
\\
\textbf{Theorem 1:} If Assumptions 1 -4 holds, then as $N\rightarrow \infty$ and $M\rightarrow\infty$, 
\begin{equation*}
\bigg| \frac{1}{K}\sum_{i=1}^K\pi_i(\theta-\bar{\theta}+\hat{\theta}_i|\mathcal{X}_i) - \pi(\theta|\mathcal{X})\bigg|_{TV} \rightarrow 0
\end{equation*}
\begin{equation*}
\bigg| \frac{1}{K}\sum_{i=1}^K\pi_i(\theta-\bar{\theta}^*+\theta^*_i|\mathcal{X}_i) - \pi(\theta|\mathcal{X})\bigg|_{TV} \rightarrow 0
\end{equation*}
The proof of Theorem 1 can be found in Appendix. \\
\\
\noindent \textbf{Remark 1:} For the center $\bar{\theta}^*$, we have $\bar{\theta}^* - \hat{\theta} = \mathcal{O}_P(\frac{1}{\sqrt{N}})$. In order to improve its performance, we can resort to the Newton-Raphson method \cite{KAK}. Under mild conditions, Newton-Raphson method can converge quadratically. Given its impact, the Newton-Raphon method only needs to be called for a few steps.\\
\\
\textbf{Remark 2:} In Lemma 1, we can replace $I(\theta_0)$ with $I_i(\hat{\theta}_i)$. Then $\sqrt{KM}(\theta - \hat{\theta}_i)$ has the same limit distribution with $\mathcal{N}(0,I^{-1}(\hat{\theta}_i))$ as $M\rightarrow \infty$, where, $\theta\sim\pi_i(\theta|\mathcal{X}_i)$. As such, if $\theta\sim\frac{1}{K}\sum_{i=1}^K\pi_i(\theta-\bar{\theta}+\hat{\theta}_i|\mathcal{X}_i)$,
\begin{equation*}
\text{Var}(\sqrt{N}(\theta - \bar{\theta}))\asymp \frac{1}{K}\sum_{i=1}^KI^{-1}_i(\hat{\theta}_i)
\end{equation*}
Because, for each $i = 1,\cdots, K$, $I_i(\hat{\theta}_i)-I(\theta_0) \rightarrow \mathcal{N}(0, \frac{1}{M}\Sigma_1)$, as a result, with Delta method, we obtain
\begin{equation*}
\sqrt{N}\left(\frac{1}{K}\sum_{i=1}^KI^{-1}_i(\hat{\theta}_i) - I^{-1}(\theta_0)\right)\rightarrow \mathcal{N}(0, \Sigma_1)
\end{equation*}
This means that the covariance matrix of our sample converges to the true one with $\mathcal{O}_P(\frac{1}{\sqrt{N}})$. \\
\\
\noindent\textbf{Remark 3:} $M$ is a user-specified parameter, which determines the gap between $\bar{\theta}$ and $\hat{\theta}$. Actually, $M$ is not necessary $\mathcal{O}(N^{\gamma})$, as long as each $\bar{\theta}_i$ is reasonable approximation to $\theta_0$, which means CLT works well for $M$. 
\section{Numerical Experiments}
\noindent We illustrate the accuracy of our method in the first three examples and compared its performance with three other methods: Consensus Monte Carlo (CMC, \cite{ScottCMC}), Weierstrass Sampler (WS, \cite{Wang}), Mposterior (MP, \cite{Mposterior}) in $L_2$ distance,
\begin{equation*}
L_2(p,q) = \int_{\mathbb{R}^d} (p(x)-q(x))^2dx
\end{equation*}
where $p, q$ are two probability density functions on $\mathbb{R}^d$. Our method is denoted by AR in the table of results of $L_2$ distances.  In Example 4, we show that our method can be applied to data augmentation cases.\\
\\
\textbf{Example 1: }(Gaussian Model) In this example, the model is assumed as follows:
\begin{equation*}
X_i|\theta \sim \mathcal{N}(\mu,\sigma^2), \quad \theta = (\mu, \sigma^2)
\end{equation*}
we sampled $X_i\sim \mathcal{N}(0,10), i = 1,\cdots, 10^6$ and chose $p(\mu, \log(\sigma))\propto 1$ as the prior. The data set was split into $K$ subsets, where we set $K= 20, 50, 100$. In Table 1 we compare the performance of the four methods. 
\begin{table}[h!]
\centering
\begin{tabular}{|| c | c | c | c | c ||}
\hline
 K               & CMC                           & AR                              & WS                             &  MP\\ \hline
 20          &$1.03\times 10^{-4}$   &$1.00\times 10^{-4}$   &$1.71\times 10^{-4}$   &$1.06\times 10^{-3}$\\ \hline
 50          &$1.54\times 10^{-4}$   &$1.33\times 10^{-4}$   &$2.87\times 10^{-4}$   &$1.53\times 10^{-3}$\\ \hline
100         &$1.48\times 10^{-4}$   &$2.24\times 10^{-4}$   &$1.58\times 10^{-4}$   &$7.25\times 10^{-4}$\\ \hline
\end{tabular}
\caption{The $L_2$ distances of our method versus others for Example 1. CMC is Consensus Monte Carlo, AR is our method, WS is Weierstrass sampler and MP is Mposterior.}
\end{table}\\
\noindent\textbf{Example 2:}  (Bayesian Logistic Model) In the Bayesian logistic model, we have observations $\{(x_i, y_i)\}_{i=1}^N$, where $x_i\in\mathbb{R}^p$ and $y_i\in\{0,1\}$, and 
\begin{equation*}
\mathbb{P}(y_i=1|x_i,\theta) = \frac{1}{1+e^{-x_i^T\theta}}
\end{equation*}
We applied this model both on synthetic and on real data sets. \\
\\
\textbf{Synthetic dataset:} The dataset $\{(x_i, y_i)\}_{i=1}^N$ consists of $N=10^5$ observations and $p = 5$.  We set $\theta = (0.3, 5, -7, 2.4, -20), x_{i1} \equiv 1$ and draw $x_{ij} \sim \mathcal{U}(0,1), j = 2,\cdots,5$. We set $K = 50, 100$. In this example, we apply the Newton-Raphon method to correct the center. Because the original center of our method is quite close to the MAP, the Netwon-Raphon method converges in this example after 5 iterations. The results are shown in Table 2. \\
\\
\noindent\textbf{Real Dataset:} We consider the \textit{Covtype} dataset \cite{WangParallel}, which consists of 581,012 observations in 54 dimensions. We consider a Bayesian logistic classification using the first $p=3$ attributes only and taking $N = 5\times 10^5$ for simplicity. During the simulations, we set $K = 100$ and $K = 500$ and call the Newton-Raphon method for 5 times. The results are shown in Table 2.  
\begin{table}[h!]
\centering
\begin{tabular}{|| c | c | c | c | c ||}
\hline
 K               & CMC                           & AR                              & WS                             &  MP\\ \hline
(Synthetic)  50          &$1.08\times 10^{-2}$   &$7.56\times 10^{-3}$   &$1.03\times 10^{-2}$   &$2.34\times 10^{-1}$\\ \hline
(Synthetic) 100         &$1.78\times 10^{-2}$   &$8.83\times 10^{-3}$   &$4.71\times 10^{-2}$   &$2.54\times 10^{-1}$\\ \hline
(Real) 100          &$3.77\times 10^{-4}$   &$3.05\times 10^{-4}$   &$9.04\times 10^{-4}$   &$6.06\times 10^{-3}$\\ \hline
(Real) 500          &$7.28\times 10^{-4}$   &$3.35\times 10^{-4}$   &$2.00\times 10^{-3}$   &$-----$\\ \hline
\end{tabular}
\caption{The $L_2$ distances of our method versus others for Example 2.}
\end{table}
\\
\noindent \textbf{Example 3:} (Beta-Bernoulli Model)  In this example, the dimension of the parameter is one and the posterior has an analytical expression, which means we can use the true posterior directly, instead of MCMC approximation. We simulated 100,000 samples from Bernoulli distribution $B(p)$ and the prior is $Beta(0.01, 0.01)$. We applied our method in two cases: $p = 0.1$, corresponding to a common scenario, and $p=0.01$, corresponding to the rare event case. We simulated $10^5$ samples from posterior or subposteriors. The $L_2$ distances are shown in Table 3 and the marginal density functions are in Figure 1. Compared with the other three methods, our method is more accurate in the cases where each subset contains only small part of information compared with the whole data set.
\begin{table}[h!]
\centering
\begin{tabular}{|| c | c | c | c | c ||}
\hline
 (p,K)               & CMC                           & AR                              & WS                             &  MP\\ \hline
 (0.1,50)          &$1.13\times 10^{-4}$   &$2.88\times 10^{-5}$   &$1.35\times 10^{-4}$   &$2.66\times 10^{-4}$\\ \hline
(0.1,100)         &$2.32\times 10^{-4}$   &$3.66\times 10^{-5}$   &$3.56\times 10^{-4}$   &$8.42\times 10^{-5}$\\ \hline
 (0.001,50)      &$1.58\times 10^{-4}$   &$3.49\times 10^{-6}$   &$1.85\times 10^{-4}$   &$2.74\times 10^{-5}$\\ \hline
(0.001,100)     &$4.78\times 10^{-4}$   &$4.02\times 10^{-6}$   &$3.37\times 10^{-4}$   &$1.15\times 10^{-4}$\\ \hline
\end{tabular}
\caption{The $L_2$ distances of our method versus others for Example 3.}
\end{table}
\begin{figure}[h]
    \centering
    \includegraphics[width=1\textwidth]{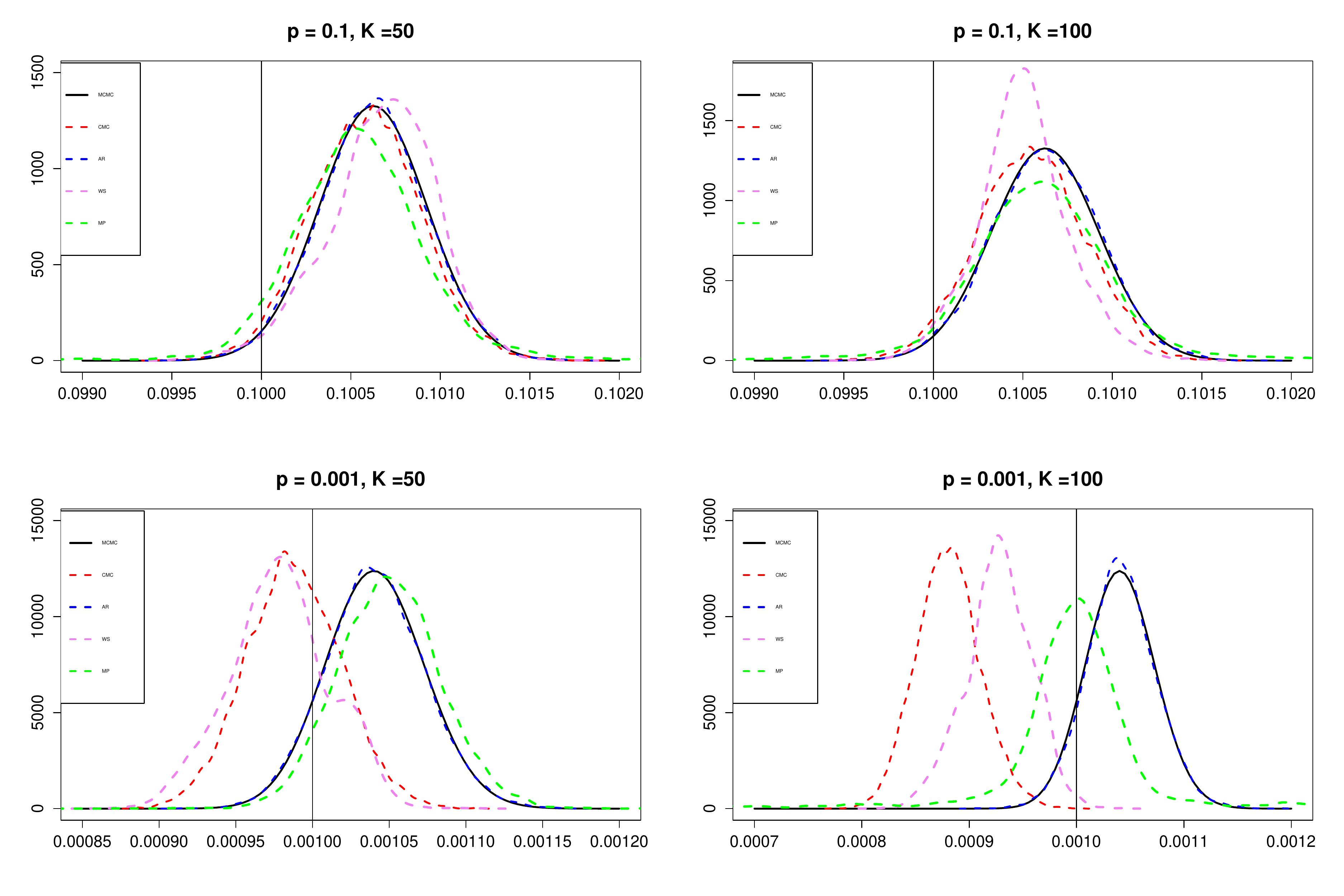}
     \caption{The graphs of probability density functions of each type of parameter. The black solid line corresponds to the true posterior, the blue dashed line is our method (AR), the red dashed line is Consensus Monte Carlo (CMC), the violet dashed line is Weierstrass sampler (WS) and the green dashed line is Mposterior (MP).}
\end{figure}
\\
\noindent\textbf{Example 4:} (Gaussian Mixture Model) In this example, we extended our method to latent variable cases. The model is
\begin{equation*}
Y_i|X_i \sim (1-p)\mathcal{N}(Y_i;\alpha X_{i1}+\beta X_{i2}, \sigma^2)+p\mathcal{N}(Y_i;0,\psi^2)
\end{equation*}
The parameter is $\theta = (\alpha, \beta, \sigma^2, \psi^2, p)$. Adding latent variables $Z_i$ and running Gibbs Sampling is the traditional way to conduct parameter inference for this model. Here, 
\begin{equation*}
\begin{split}
Y_i|Z_i=0,X_i &\sim \mathcal{N}(Y_i;\alpha X_{i1}+\beta X_{i2}, \sigma^2) \\
Y_i|Z_i=1,X_i &\sim \mathcal{N}(Y_i;0,\psi^2)
\end{split}
\end{equation*} 
The posterior of $\theta,Z$ given the observations $\{(X_i,Y_i)\}$ is 
\begin{equation*}
p(\theta,Z|Y,X)\propto \left(\prod_{i=1}^N \left(\phi(y_i|\alpha x_{i1}+\beta x_{i2}, \sigma^2)\right)^{1-Z_i} \left(\phi(y_i|0,\psi^2)\right)^{Z_i}\pi(Z_i|p)\right)\pi(\alpha)\pi(\beta)\pi(\sigma^2)\pi(\psi^2)\pi(p)
\end{equation*}
The priors are chosen to be
\begin{equation*}
\begin{split}
\alpha \sim \mathcal{N}(m_{\alpha}, \sigma^2_{\alpha}),\quad \beta \sim \mathcal{N}(m_{\beta}, \sigma^2_{\beta}),\\
\sigma^2 \sim \mathcal{IG}(\alpha_{\sigma}, \beta_{\sigma}),\quad \psi^2 \sim \mathcal{IG}(\alpha_{\psi}, \beta_{\psi}),\\
Z_i|p \sim Bernoulli(p),\quad p \sim Beta(\lambda, \eta).\\
\end{split}
\end{equation*}
These priors are conjugate for the model. That is, 
\begin{equation*}
z_i|X_i, Y_i, \theta \sim Bernoulli(p_i^*),\quad p_i^* = \frac{p\phi(y_i|0,\psi^2)}{(1-p)\phi(y_i|\alpha x_{i1}+\beta x_{i2}, \sigma^2)+p\phi(y_i|0,\psi^2)}
\end{equation*}
\begin{equation*}
\alpha|X,Y,Z,\theta_{-\alpha} \sim \mathcal{N}(m_{\alpha}^*, {\sigma_{\alpha}^*}^2), \quad m_{\alpha}^* = \frac{\sum_{i=1}^N\frac{(y_i-\beta x_{i2})x_{i1}(1-z_i)}{\sigma^2}+\frac{m_{\alpha}}{\sigma_{\alpha}^2}}{\sum_{i=1}^N\frac{x_{i1}^2(1-z_i)}{\sigma^2}+\frac{1}{\sigma_{\alpha}^2}}\quad {\sigma_{\alpha}^*}^2 = \frac{1}{\sum_{i=1}^N\frac{x_{i1}^2(1-z_i)}{\sigma^2}+\frac{1}{\sigma_{\alpha}^2}}
\end{equation*}
\begin{equation*}
\beta|X,Y,Z,\theta_{-\beta} \sim \mathcal{N}(m_{\beta}^*, {\sigma_{\beta}^*}^2), \quad m_{\beta}^* = \frac{\sum_{i=1}^N\frac{(y_i-\alpha x_{i1})x_{i2}(1-z_i)}{\sigma^2}+\frac{m_{\beta}}{\sigma_{\beta}^2}}{\sum_{i=1}^N\frac{x_{i2}^2(1-z_i)}{\sigma^2}+\frac{1}{\sigma_{\beta}^2}}\quad {\sigma_{\beta}^*}^2 = \frac{1}{\sum_{i=1}^N\frac{x_{i2}^2(1-z_i)}{\sigma^2}+\frac{1}{\sigma_{\beta}^2}}
\end{equation*}
\begin{equation*}
\sigma^2|X,Y,Z,\theta_{-\sigma^2} \sim \mathcal{IG}(\alpha_{\sigma}^*,\beta_{\sigma}^*),\quad \alpha_{\sigma}^* = \alpha_{\sigma}+\frac{1}{2}\sum_{i=1}^N(1-z_i), \quad \beta_{\sigma}^* = \beta_{\sigma} + \frac{1}{2}\sum_{i=1}^N(1-z_i)(\alpha x_{i1} + \beta x_{i2} -y_i)^2
\end{equation*}
\begin{equation*}
\psi^2|X,Y,Z,\theta_{-\psi^2} \sim \mathcal{IG}(\alpha_{\psi}^*,\beta_{\psi}^*),\quad \alpha_{\psi}^* = \alpha_{\psi}+\frac{1}{2}\sum_{i=1}^Nz_i, \quad \beta_{\psi}^* = \beta_{\psi} + \frac{1}{2}\sum_{i=1}^Nz_iy_i^2
\end{equation*}
\begin{equation*}
p|X,Y,Z,\theta_{-p} \sim Beta\left(\lambda+\sum_{i=1}^Nz_i, \eta+\sum_{i=1}^N(1-z_i)\right)
\end{equation*}
Even though this model is conjugate, simulating a label $Z_i$ for each data point $(x_i, y_i)$ at each iteration is too expensive to use the Gibbs Sampling in big data context. In our simulation, we set $\theta = (2, 5, 1, 10, 0.05)$, $N = 10^6$ and $K = 50, M = N/K$.  For each subset $\mathcal{X}_i = \{x_{i1},\cdots,x_{iM}\}$, imagine that we have a sequence of artificial observations $\{(x_i^*, y_i^*)\}_{i=1}^N$ by repeating $\mathcal{X}_{i}$ with $K$ times, that is,
\begin{equation*}
x_t^* = x_{ij}, \qquad\text{for} \quad (j-1)K+1\leq t \leq jK
\end{equation*}
For each $x_{ij}$, it appears $K$ times, which means its corresponding labels follow binomial distribution.  The Gibbs updating procedure should be changed:
\begin{equation*}
z_{ij}|x_{ij},y_{ij},\theta =Bernoulli(K, p_{ij}^*),\quad p_{ij}^*=\frac{p\phi(y_{ij}|0,\psi^2)}{(1-p)\phi(y_{ij}|\alpha x_{{ij}1}+\beta x_{{ij}2}, \sigma^2)+p\phi(y_{ij}|0,\psi^2)}
\end{equation*}
\begin{equation*}
\alpha|\mathcal{X}_i,\mathcal{Z}_i,\theta_{-\alpha} \sim \mathcal{N}(m_{\alpha}^*, {\sigma_{\alpha}^*}^2), \quad m_{\alpha}^* = \frac{\sum_{j=1}^M\frac{(y_{ij}-\beta x_{{ij}2})x_{{ij}1}(K-z_{ij})}{\sigma^2}+\frac{m_{\alpha}}{\sigma_{\alpha}^2}}{\sum_{j=1}^M\frac{x_{{ij}1}^2(K-z_{ij})}{\sigma^2}+\frac{1}{\sigma_{\alpha}^2}}\quad {\sigma_{\alpha}^*}^2 = \frac{1}{\sum_{j=1}^M\frac{x_{{ij}1}^2(K-z_{ij})}{\sigma^2}+\frac{1}{\sigma_{\alpha}^2}}
\end{equation*}
\begin{equation*}
\beta|\mathcal{X}_i,\mathcal{Z}_i,\theta_{-\beta} \sim \mathcal{N}(m_{\beta}^*, {\sigma_{\beta}^*}^2), \quad m_{\beta}^* = \frac{\sum_{j=1}^M\frac{(y_{ij}-\alpha x_{{ij}1})x_{{ij}2}(K-z_{ij})}{\sigma^2}+\frac{m_{\beta}}{\sigma_{\beta}^2}}{\sum_{j=1}^M\frac{x_{{ij}2}^2(K-z_{ij})}{\sigma^2}+\frac{1}{\sigma_{\beta}^2}}\quad {\sigma_{\beta}^*}^2 = \frac{1}{\sum_{j=1}^M\frac{x_{{ij}2}^2(K-z_{ij})}{\sigma^2}+\frac{1}{\sigma_{\beta}^2}}
\end{equation*}
\begin{equation*}
\sigma^2|\mathcal{X}_i,\mathcal{Z}_i,\theta_{-\sigma^2} \sim \mathcal{IG}(\alpha_{\sigma}^*,\beta_{\sigma}^*),\quad \alpha_{\sigma}^* = \alpha_{\sigma}+\frac{1}{2}\sum_{j=1}^M(K-z_{ij}), \quad \beta_{\sigma}^* = \beta_{\sigma} + \frac{1}{2}\sum_{j=1}^M(K-z_{ij})(\alpha x_{{ij}1} + \beta x_{{ij}2} -y_{ij})^2
\end{equation*}
\begin{equation*}
\psi^2|\mathcal{X}_i,\mathcal{Z}_i,\theta_{-\psi^2} \sim \mathcal{IG}(\alpha_{\psi}^*,\beta_{\psi}^*),\quad \alpha_{\psi}^* = \alpha_{\psi}+\frac{1}{2}\sum_{j=1}^Mz_{ij}, \quad \beta_{\psi}^* = \beta_{\psi} + \frac{1}{2}\sum_{j=1}^Mz_{ij}y_{ij}^2
\end{equation*}
\begin{equation*}
p|\mathcal{X}_i,\mathcal{Z}_i,\theta_{-p} \sim Beta\left(\lambda+\sum_{j=1}^Mz_{ij}, \eta+\sum_{j=1}^M(K-z_{ij})\right)
\end{equation*}
In this example, the Figure 2 shows that our method is quite appealing in accuracy. 
\begin{figure}[!h]
    \centering
    \includegraphics[width=0.6\textwidth]{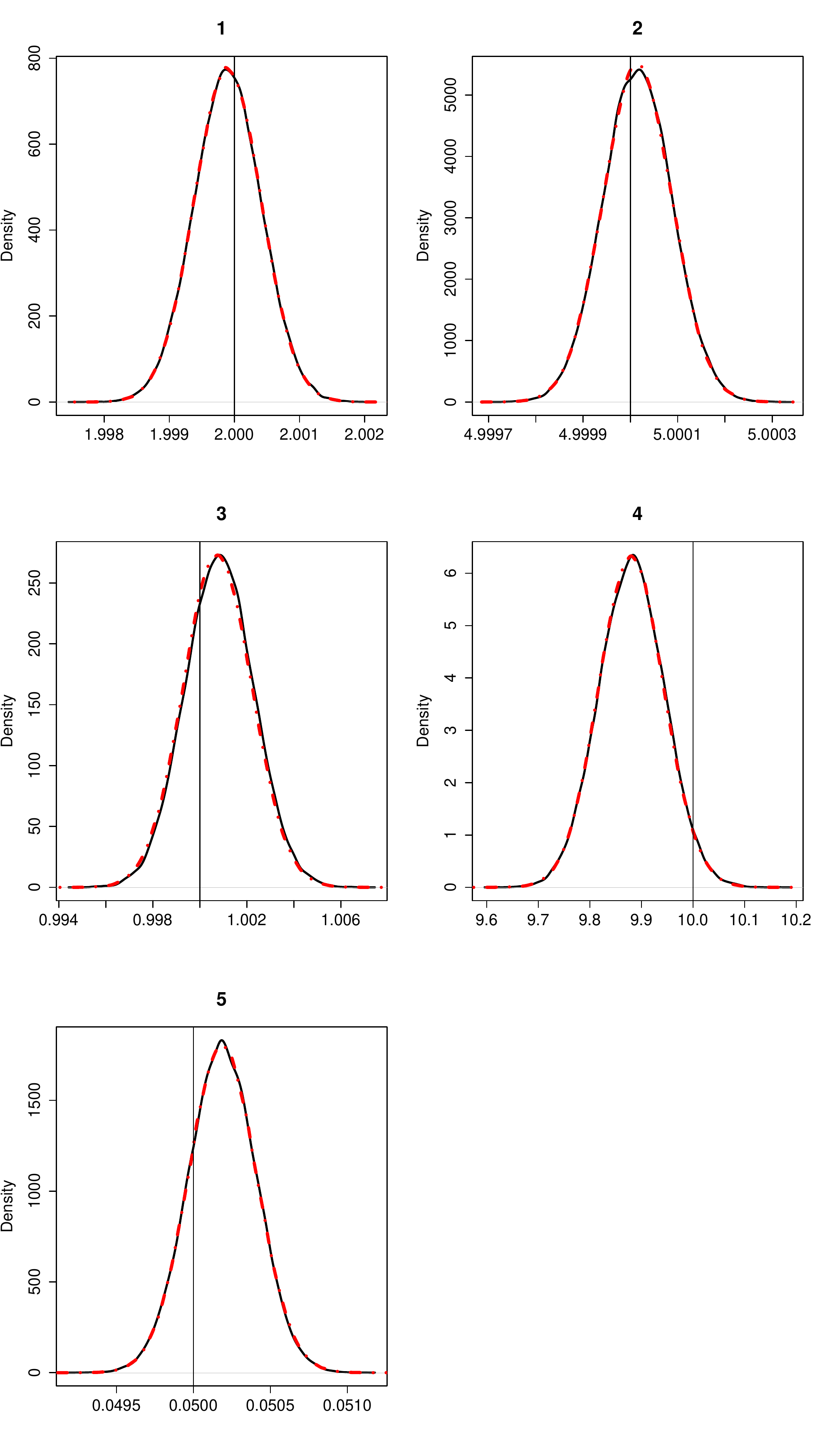}
    \caption{The graphs of probability density functions of each type of parameter. Vertical lines mark the parameter values from which the data set is generated. Black solid lines represent the marginal pdf of samples from full MCMC while red dashed lines represent the ones from our method}
    \end{figure}
\section{Conclusion}
\noindent In this article, we proposed a new combination of samples from rescaled subposteriors to approximate the overall posterior and gave its mathematical justification. In order to show its validation in practice, we applied it on several common models. Compared with classical parallel approaches, our method is more reasonable at a statistical level, shares the same computation cost in parallel stage and the combination stage is very cheap, without the necessity of running an additional MCMC. At the same time, according to the simulations, our method is quite accurate and satisfactory.
\\
\\

\section{Appendix}
\noindent In this section, we give a proof of Theorem 1. First, let's introduce some useful lemmas. \\
\\
\textbf{Lemma 1:}  The distribution of $\sqrt{N}(\theta-\hat{\theta}_i)$, where $\theta\sim\pi_i(\theta|\mathcal{X}_i)$ converges to the normal distribution $\mathcal{N}(\mathbf{0},I^{-1}(\theta_0))$ under the total variation metric, i.e.
\begin{equation}
\int_{\mathcal{R}^d}\bigg|\pi_i(\theta|\mathcal{X}_i)-\phi(\theta;\hat{\theta}_i,I^{-1}(\theta_0))\bigg|d\theta \rightarrow 0 
\end{equation}
\textbf{Proof:} Denote $t = \sqrt{N}(\theta - \hat{\theta}_i)$, $\theta \sim \pi_i(\theta|\mathcal{X}_i)$, then 
\begin{equation*}
t \sim C_i^{-1}\pi(\hat{\theta}_i+\frac{t}{\sqrt{N}})\exp\left[K\ell_i(\hat{\theta}_i+\frac{t}{\sqrt{N}})- K\ell_i(\hat{\theta}_i)\right]
\end{equation*}
where $C_i = \int_{\mathbb{R}^d}\pi(\hat{\theta}_i+\frac{t}{\sqrt{N}})\exp\left[K\ell_i(\hat{\theta}_i+\frac{t}{\sqrt{N}})- K\ell_i(\hat{\theta}_i)\right]dt$. Denote
\begin{equation*}
g_i(t) = \pi(\hat{\theta}_i+\frac{t}{\sqrt{N}})\exp\left[K\ell_i(\hat{\theta}_i+\frac{t}{\sqrt{N}})- K\ell_i(\hat{\theta}_i)\right] - \pi(\theta_0)\exp\left[-\frac{t^TI(\theta_0)t}{2}\right]
\end{equation*}
On $A_1 = \{t: ||t||>\delta_0\sqrt{N}\}$, we have 
\begin{equation*}
K\ell_i(\hat{\theta}_i+\frac{t}{\sqrt{N}})- K\ell_i(\hat{\theta}_i) < -N\epsilon
\end{equation*}
Then, 
\begin{equation*}
\int_{A_1}g_i(t)dt \rightarrow 0
\end{equation*}
On $A_2 = \{t: ||t||\leq\delta_0\sqrt{N}\}$, we have 
\begin{equation*}
K\ell_i(\hat{\theta}_i+\frac{t}{\sqrt{N}})- K\ell_i(\hat{\theta}_i) = -\frac{1}{2}t^T\hat{I}_it + R_i(t)
\end{equation*}
where $R_i(t) = \frac{K}{6}(\frac{1}{\sqrt{N}})^3\sum_{j=1}^M\sum_{p=1}^d\sum_{q=1}^d\sum_{r=1}^dt_pt_qt_r\frac{\partial^3}{\partial\theta_p\partial\theta_q\partial\theta_r}\log{f(x_{ij}|\theta')}$, $\theta'$ lies in the line segment between $\hat{\theta}_i$ and $\hat{\theta}_i+\frac{t}{\sqrt{N}}$, and 
\begin{equation*}
(\hat{I}_i)_{pq} = \frac{1}{M}\sum_{j=1}^M\left[-\frac{\partial^2}{\partial \theta_p\partial \theta_q}\log{f(x_{ij}|\theta)}\right]\bigg|_{\hat{\theta}_i}
\end{equation*}
For each $t\in A_2$, we obtain $R_i(t)\rightarrow 0$ and $ \hat{I}_i\rightarrow I(\theta_0)$ as $M\rightarrow \infty$. Hence, $g_i(t)\rightarrow 0$.  Besides,
\begin{equation*}
|R_i(t)| \leq \frac{1}{6}\delta_0\frac{t^2}{N}d^3K\sum_{j=1}^MM(x_{ij})\leq \frac{1}{4}t^T\hat{I}_it
\end{equation*}
As a result,
\begin{equation*}
\exp\left[K\ell_i(\hat{\theta}_i+\frac{t}{\sqrt{N}})- K\ell_i(\hat{\theta}_i)\right]\leq \exp\left[-\frac{1}{4}t^T\hat{I}_it\right]\leq \exp\left[-\frac{t^TI(\theta_0)t}{8}\right]
\end{equation*}
Therefore, $|g_i(t)|$ is dominated by an integrable function on $A_2$. Thus, we obtain $\int_{\mathbb{R}^d}|g_i(t)|dt \rightarrow 0$ as $M\rightarrow \infty$. Hence, we have immediately
\begin{equation*}
C_i\rightarrow \pi(\theta_0)\sqrt{(2\pi)^d/\det{I(\theta_0)}}
\end{equation*}
\begin{equation*}
\begin{split}
&\int \bigg|C_i^{-1}\pi(\hat{\theta}_i+\frac{t}{\sqrt{N}})\exp\left[K\ell_i(\hat{\theta}_i+\frac{t}{\sqrt{N}})- K\ell_i(\hat{\theta}_i)\right] - \sqrt{\frac{\det{I(\theta_0)}}{(2\pi)^d}}\exp\left[-\frac{1}{2}t^TI(\theta_0)t\right]\bigg|dt\\
&\leq C_i^{-1}\int |g_i(t)|dt + \int \bigg| C_i^{-1}\pi(\theta_0)\exp\left[-\frac{1}{2}t^TI(\theta_0)t\right]- \sqrt{\frac{\det{I(\theta_0)}}{(2\pi)^d}}\exp\left[-\frac{1}{2}t^TI(\theta_0)t\right] \bigg|dt\\
&\leq C_i^{-1}\int |g_i(t)|dt + \bigg| C_i^{-1}\pi(\theta_0)/\sqrt{\frac{\det{I(\theta_0)}}{(2\pi)^d}}-1\bigg| \rightarrow 0
\end{split}
\end{equation*}
\textbf{Corollary 1:} Denote $\bar{\theta} = \frac{1}{K}\sum_{i=1}^K\hat{\theta}_i$. Then 
\begin{equation*}
\int_{\mathcal{R}^d}\bigg|\frac{1}{K}\sum_{i=1}^K\pi_i(\theta-\bar{\theta}+\hat{\theta}_i|\mathcal{X}_i)-\phi(\theta;\bar{\theta},I^{-1}(\theta_0))\bigg|d\theta \rightarrow 0
\end{equation*}
\textbf{Corollary 2:} Denote $\theta^*_i = \int \theta\pi_i(\theta|\mathcal{X}_i)d\theta$, then $\sqrt{N}(\theta^*_i - \hat{\theta}_i)\rightarrow 0$.\\
\\
\textbf{Proof:} Proceeding as in the proof of Lemma 1 and using the assumption of finite expectation of the prior, we can have
\begin{equation*}
\int |t| \bigg|C_i^{-1}\pi(\hat{\theta}_i+\frac{t}{\sqrt{N}})\exp\left[K\ell_i(\hat{\theta}_i+\frac{t}{\sqrt{N}})- K\ell_i(\hat{\theta}_i)\right] - \sqrt{\frac{\det{I(\theta_0)}}{(2\pi)^d}}\exp\left[-\frac{1}{2}t^TI(\theta_0)t\right]\bigg|dt\rightarrow 0
\end{equation*}
This implies 
\begin{equation*}
\int tC_i^{-1}\pi(\hat{\theta}_i+\frac{t}{\sqrt{N}})\exp\left[K\ell_i(\hat{\theta}_i+\frac{t}{\sqrt{N}})- K\ell_i(\hat{\theta}_i)\right]dt \rightarrow \int t\sqrt{\frac{\det{I(\theta_0)}}{(2\pi)^d}}\exp\left[-\frac{1}{2}t^TI(\theta_0)t\right]dt =0
\end{equation*}
Therefore, 
\begin{equation*}
\sqrt{N}(\theta^*_i - \hat{\theta}_i) = \int tC_i^{-1}\pi(\hat{\theta}_i+\frac{t}{\sqrt{N}})\exp\left[K\ell_i(\hat{\theta}_i+\frac{t}{\sqrt{N}})- K\ell_i(\hat{\theta}_i)\right]dt \rightarrow 0
\end{equation*}
\textbf{Lemma 2:} For two multivariate normal distributions $P = \mathcal{N}(\mu_1,\Sigma)$ and $Q = \mathcal{N}(\mu_2,\Sigma)$, their Kullback-Leibler divergence is 
\begin{equation*}
KL(P|Q) = KL(Q|P) = \frac{1}{2}(\mu_1-\mu_2)^T\Sigma^{-1}(\mu_1-\mu_2)
\end{equation*}
\textbf{Lemma 3:} For two probability measures $P$ and $Q$, we have following inequality
\begin{equation*}
\bigg| P - Q \bigg|_{TV}\leq 2\sqrt{KL(P|Q)}
\end{equation*}
\textbf{Lemma 4:} For each $i \in \{1, \cdots, K\}$, we have $\hat{\theta}_i-\theta_0\rightarrow \mathcal{N}(0, \frac{1}{M}I^{-1}(\theta_0))$, then $\bar{\theta}-\theta_0\rightarrow \mathcal{N}(0, \frac{1}{N}I^{-1}(\theta_0))$ and $\bar{\theta} - \hat{\theta} = O_p(\frac{1}{\sqrt{N}})$. \\
\\
 \textbf{Proof:} Based on the above lemmas, we have 
 \begin{equation*}
 \sqrt{N}(\bar{\theta}-\theta_0) = \frac{1}{\sqrt{K}}\sum_{i=1}^K\sqrt{M}(\hat{\theta}_i-\theta_0)\rightarrow \mathcal{N}(0,I^{-1}(\theta_0))
 \end{equation*}
 \begin{equation*}
 ||\bar{\theta}-\hat{\theta}||\leq ||\bar{\theta}-\theta_0||+||\hat{\theta}-\theta_0|| = O_P(\frac{1}{\sqrt{N}})
 \end{equation*}
\textbf{Theorem 1:} If the Assumptions 1 -4 holds, then as $N\rightarrow \infty$ and $M\rightarrow\infty$, 
\begin{equation*}
\bigg| \frac{1}{K}\sum_{i=1}^K\pi_i(\theta-\bar{\theta}+\hat{\theta}_i|\mathcal{X}_i) - \pi(\theta|\mathcal{X})\bigg|_{TV} \rightarrow 0
\end{equation*}
\begin{equation*}
\bigg| \frac{1}{K}\sum_{i=1}^K\pi_i(\theta-\bar{\theta}^*+\theta^*_i|\mathcal{X}_i) - \pi(\theta|\mathcal{X})\bigg|_{TV} \rightarrow 0
\end{equation*}
\textbf{Proof:}
\begin{equation*}
\begin{split}
&\bigg| \frac{1}{K}\sum_{i=1}^K\pi_i(\theta-\bar{\theta}+\hat{\theta}_i|\mathcal{X}_i) - \pi(\theta|\mathcal{X})\bigg|_{TV}\\
&\leq\frac{1}{K}\sum_{i=1}^K\bigg|\pi_i(\theta-\bar{\theta}+\hat{\theta}_i|\mathcal{X}_i) - \mathcal{N}(\theta;\bar{\theta},\frac{1}{\sqrt{N}}I^{-1}(\theta_0))\bigg|_{TV} \\
&\quad+ \bigg|\mathcal{N}(\theta;\bar{\theta},\frac{1}{\sqrt{N}}I^{-1}(\theta_0)) - \mathcal{N}(\theta;\hat{\theta},\frac{1}{\sqrt{N}}I^{-1}(\theta_0))\bigg|_{TV}\\
&\quad+\bigg|\mathcal{N}(\theta;\hat{\theta},\frac{1}{\sqrt{N}}I^{-1}(\theta_0)) - \pi(\theta|\mathcal{X})\bigg|_{TV}\\
&\rightarrow 0
\end{split}
\end{equation*}
\begin{equation*}
\begin{split}
&\bigg|\pi(\theta-\bar{\theta}^*+\theta^*_i|\mathcal{X}_i) - \pi_i(\theta-\bar{\theta}+\hat{\theta}_i|\mathcal{X}_i)\bigg|_{TV}\\
&\leq\bigg|\pi(\theta-\bar{\theta}^*+\theta^*_i|\mathcal{X}_i) - \mathcal{N}(\theta;\bar{\theta}^*-\theta_i^*+\hat{\theta}_i, \frac{1}{N}I^{-1}(\theta_0))\bigg|_{TV}\\
&+\bigg|\mathcal{N}(\theta;\bar{\theta}^*-\theta_i^*+\hat{\theta}_i, \frac{1}{N}I^{-1}(\theta_0))-\mathcal{N}(\theta;\bar{\theta}, \frac{1}{N}I^{-1}(\theta_0))\bigg|_{TV}\\
&+\bigg|\mathcal{N}(\theta;\bar{\theta}, \frac{1}{N}I^{-1}(\theta_0))-\pi_i(\theta-\bar{\theta}+\hat{\theta}_i|\mathcal{X}_i)\bigg|_{TV}\\
&\rightarrow 0
\end{split}
\end{equation*}
\begin{equation*}
\begin{split}
&\bigg| \frac{1}{K}\sum_{i=1}^K\pi_i(\theta-\bar{\theta}^*+\theta^*_i|\mathcal{X}_i) - \pi(\theta|\mathcal{X})\bigg|_{TV}\\
&\leq\frac{1}{K}\sum_{i=1}^K\bigg|\pi_i(\theta-\bar{\theta}^*+\theta^*_i|\mathcal{X}_i) - \pi_i(\theta-\bar{\theta}+\hat{\theta}_i|\mathcal{X}_i)\bigg|_{TV}\\
&+\bigg| \frac{1}{K}\sum_{i=1}^K\pi_i(\theta-\bar{\theta}+\hat{\theta}_i|\mathcal{X}_i) - \pi(\theta|\mathcal{X})\bigg|_{TV}\\
&\rightarrow 0
\end{split}
\end{equation*}
\end{document}